\documentstyle[12pt,a4]{article}
\renewcommand{\baselinestretch}{1.6}
\begin{document}
\title{Analysis by neutron activation analysis a some ancient ceramics from
Romanian territories}
\author{Agata Olariu\\
{\em National Institute for Physics and Nuclear Engineering}\\
{\em P.O.Box MG-6, 76900 Bucharest Magurele, Romania}\\}

\maketitle

\section{Introduction}

Ceramics is the most common archaeological material and therefore it is 
very used material by the historians to draw temporal and cultural
characterisations. The importance of knowledge the the compositional scheme of 
the pottery is well known$^{1-5}$ although very rarely one can get important
conclusions from the elemental analysis of the potsherds$^{6-8}$.

In this paper we have analyzed samples of Neolthic ceramics from Cucuteni
Scanteia - Vaslui county (History Museum from Iasi), and Neolithic and Dacian 
ceramics 
from Magurele - near Bucharest (collection of the Scholl from Magurele), 
Romania,  by the method of neutron activation analysis (NAA).

In Table 1 it is presented the list of the analyzed ceramics samples.

\begin{table}[h]
\newlabel{}
\caption{{\bf Table 1}. List of archaeological ceramic objects from Cucuteni 
and Magurele analyzed by neutron activation analysis}\\
\begin{center}
\begin{tabular}[h]{ccc}
\hline
Sample  & Object & Provenance\\
\hline
\hline\\
CS1, CS2, ..., CS17   & ceramic sherds & Cucuteni Sc\^{a}nteia, county Vaslui\\
CS18, CS19             & pigments, clay ball & Cucuteni Sc\^{a}nteia\\
M1dac, M2dac, M3dac      & ceramic sherds & Dacian, Magurele \\
M4neo, M5neo, M6neo             & ceramic sherds & Neolithic Magurele\\
\end{tabular}
\end{center}
\end{table}
\newpage
\section{Experimental method}

The samples listed in the Table 1 have been analyzed by neutron activation
analysis. 
We considered that the analysis should give an image of the bulk of the objects
and therefore the surface of the shards was removed. Also we had into
consideration the homogeinity of the samples and that the samples must 
be representative for the whole object.
Samples of 10-30 mg of potsherds have been cut, weighted and wrapped 
individually in plastic foil. The samples of 10-30 mg have been irradiated at
at the rabbit system of VVR-S reactor of the NIPNE,
Bucharest-Magurele, at a flux of 1.5 x 10$^{12}$ 
neutrons/cm$^2$$\cdot$sec$^{-1}$, for a period of 30 minutes. A standard 
spectroscopic pure metallic
copper was used as neutron flux standard. The radioactivity of the samples has
been measured after a decay time of 1...20 h, and again after a decay time of 
10...14 d. The measurements have been performed
with a Ge(Li) detector, 135 cm$^3$ coupled at PC with a MCA interface. The
system gave a resolution of 2.4 keV at 1.33 MeV ($^{60}$Co). 
The following elements have been noticed: Fe, K, La, Mn, Na, Sc and Sm.\\ 

\section{Results and discussions}

In the Table 2 are given the results of the neutron activation analysis
on the ancient potsherds from Cucuteni Scanteia and Magurele.
The concentrations are given in ppm, and when the concentration was larger than
10,000 ppm the result was given in percents. The considered statistical errors 
have been for La, Mn, Na, Sc and Sm $<$5\% and $\approx$10\%
for K and Fe. 
The diagram from the Fig. 1 describes the analyzed potsherds from the point 
of view of all considered elements in the samples. The minimum and maximum 
limits of the concentrations are shown for all observed elements. One can 
notice that
the 2 compositional schemes are not completely separated and there are regions
of interference between Cucuteni group of the objects and Magurele group.\\
Searching an characteristic element or ratio of elements for a given region,
we have noticed that the ratios Na/Mn, La/Sc and La/Sm
could be considered  representative and characteristic for a group of analyzed 
sherds.
In Fig. 2a and Fig. 2b are shown the diagrams of the ratios of
concentrations Na/Mn versus La/Sc, and respectively Na/Mn versus La/Sm,
and one could observe a relative synchronozation of the analyzed objects 
on the provenance and culture on the mentioned ratios.

\newpage
\small
\begin{table}
\newlabel{}
\caption{{\bf Table 2}. Concentrations, ancient ceramics from 
Cucuteni-Scanteia 
and Magurele, by NAA. The results are given in ppm, and for concentrations
larger than 10,000 ppm, the results are given in \%}\\

\begin{tabular}{lccccccc}
\hline
Sample &Manganese &Natrium & Potasium & Samarium & Lantanum & Scandium & Iron\\
\hline
\hline

CS1	&       828	&	8450	&	3.1\%	&	8.3	& 40	&	21	&	6.2\%\\
CS2	&	834	&	6480	&	2.8\%	&	7	&	39	&	18	&	7.0\%	\\
CS3	&	840	&	5790	&	2.2\%	&	7.2	&	39	&	18	&	4.7\%	\\
CS4	&	728	&	7540	&	2.4\%	&	9	&	36	&	17	&	3.9\%	\\
CS5	&	1030	&	9360	&	2.4\%	&	10	&	37	&	17	&	5.9\%	\\
CS6	&	905	&	8380	&	3.3\%	&	11	&	55	&	23	&	6.1\%	\\
CS7	&	1070	&	1.28\%	&	4.1\%	&	11	&	42	&	19	&	7.9\%	\\
CS8	&	375	&	7780	&	2.9\%	&	8.2	&	42	&	18	&	5.1\%	\\
CS9	&	985	&	7460	&	3.0\%	&	9.5	&	40	&	23	&	5.5\%	\\
CS10	&	764	&	9830	&	4.0\%	&	10	&	31	&	20	&	7.7\&	\\
CS11	&	905	&	9430	&	3.4\%	&	11.5	&	53	&	24	&	6.8\%	\\
CS12	&	1130	&	5880	&	3.1\%	&	9.5	&	42	&	20	&	6.0\%	\\
CS13	&	1040	&	7310	&	3.3\%	&	13	&	50	&	22	&	7.1\%	\\
CS14	&	1180	&	1.17\%	&	3.2\%	&	11	&	40	&	25	&	5.0\%	\\
CS15	&	608	&	4880	&	3.6\%	&	9.3	&	43	&	26	&	7.6\%	\\
CS16	&	634	&	7050	&	3.2\%	&	6	&	34	&	24	&	8.4\%	\\
CS17	&	607	&	8560	&	3.3\%	&	10.5	&	44	&	19	&	6.2\%	\\
CS18	&	2460	&	5980	&	2.6\%	&	7	&	30	&	18	&	31\%	\\
CS19	&	1.92\%	&	5890	&	9050	&	4	&	35	&	6	&	3.5\%	\\
M1dac	&	873	&	1.59\%	&	3.4\%	&	13	&	60	&	23	&	6.0\%	\\
M2dac	&	697	&	1.61\%	&	4.1\%	&	15	&	70	&	21	&	6.2\%	\\
M3dac	&	955	&	1.5\%	&	3.2\%	&	13	&	60	&	22	&	9.3\%	\\
M4neo	&	479	&	1.4\%	&	2.1\%	&	9	&	34	&	14	&	5.0\%	\\
M5neo	&	583	&	1.5\%	&	3.5\%	&	11	&	44	&	19	&	7.2\%	\\
M6neo	&	405	&	1.4\%	&	2.2\%	&	9	&	38	&	15	&	6.4\%	\\

\end{tabular}
\end{table}
\noindent
{\bf\large References}

\noindent
1. A. Aspinal, D. N. Slater, "Neutron activation analysis of medieval 
ceramics", Nature 217 (1968) 368\\
2. J. S. Olin and Ed. V. Sayre, "Trace analysis of English and American pottery
of the american colonial period" The 1968 Intern. Conference of Modern Trends
in Activation Analysis" (1968) p. 207\\ 
3. N. Saleh, A. Hallak and C. Bennet, "PIXE analysis of ancient ordanian
pottery", Nuclear Instruments and Methods 181 (1981) p. 527\\
4. Ed. Sayre, "Activation Analysis applications in art and archaeology", in
Advancees in Activation Analysis, eds. J.M.A. Lenihan, S.J. Thomson and V.P.
Guinn, Academic Press, London, p.157\\
5. Ch. Lahanier, F.D. Preusser and L. Van Zelst, "Study and conservation of
museum objects: use of classical analytical techniques", Nuclear Instruments
and Methos, B14 (1986) p.2\\
6. Zvi Goffer, Archaeological Chemistry, Chemical Analysis, Vol. 55, eds. P.J.
Elving J. D. Winefordner, John Wiley \& Sons, p.108\\
7. I. Perlman and F. Assaro, "Deduction of provenience of pottery from trace
element analysis", Scientific Methods in Medieval Archaeology, ed. R. Berger
Univ. of California Press (1970) p.389\\
8. A. Millet and H. Catling, "Composition and provenance: a challenge",
Archaeometry, Vol. 9 (1966) p.92\\

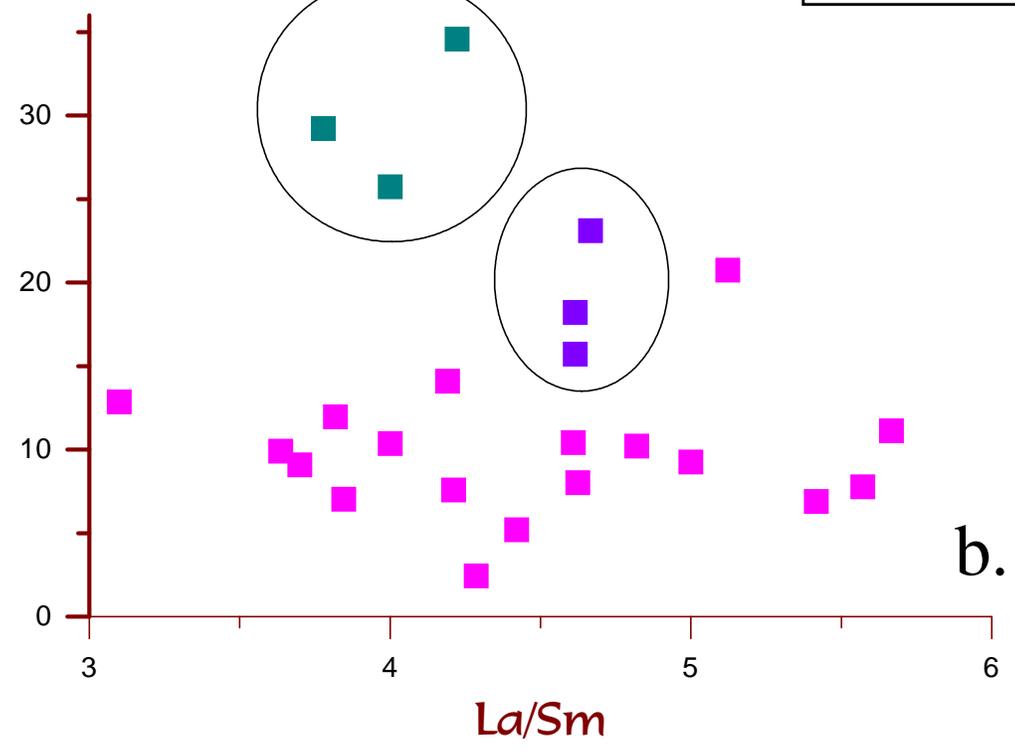
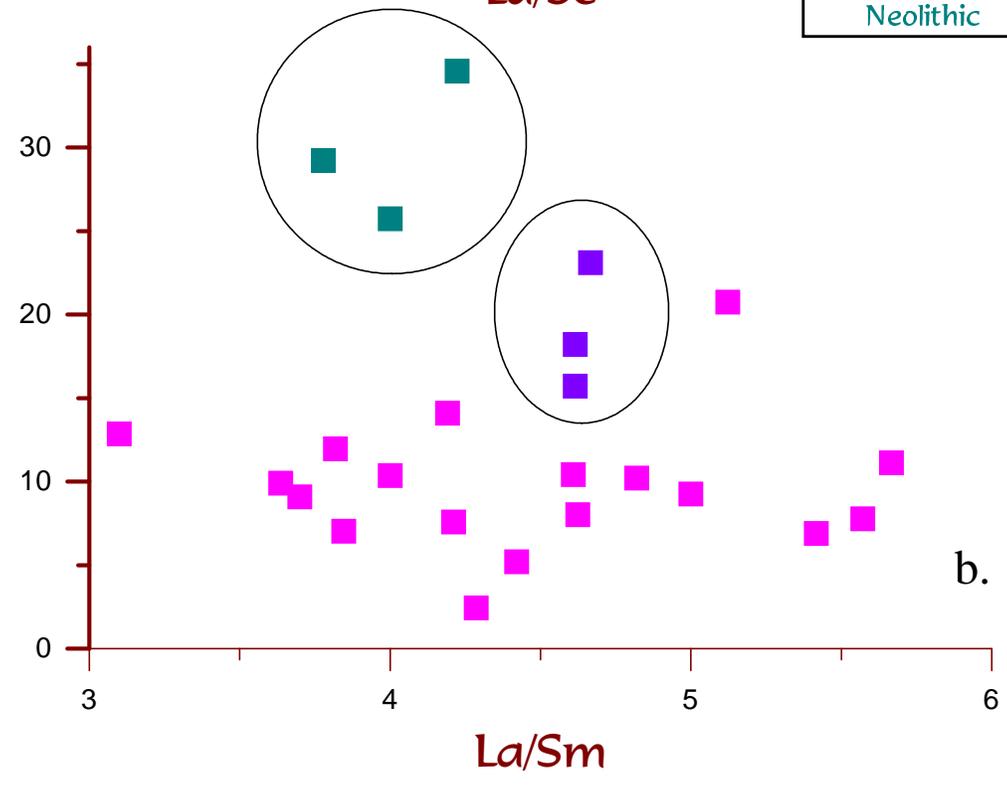
\end{document}